\documentclass[usenatbib,conference]{basi}
\usepackage[T1]{fontenc}
\usepackage[british]{babel}
\usepackage[varg]{txfonts}
\usepackage{rotating}
\usepackage{dcolumn}

\usepackage[breaklinks=false,backref=false]{hyperref}
\hypersetup{
    bookmarksopen=false,
    bookmarksnumbered=false,
    unicode=true,           
    pdftoolbar=true,        
    pdfmenubar=true,        
    pdffitwindow=false,     
    pdfstartview={FitH},    
    pdfnewwindow=true,      
    colorlinks=true,       
    linkcolor=red,          
    citecolor=magenta,        
    filecolor=green,      
    urlcolor=blue           
}

\begin{document}
\title[iSpec: An integrated software framework for the analysis of stellar spectra]{iSpec: An integrated software framework for the analysis of stellar spectra}
%
\author[S. Blanco-Cuaresma et al.]{ S. Blanco-Cuaresma$^{1, 2}$,
        C. Soubiran$^{1, 2}$,
        P. Jofr\'e$^{1, 2, 3}$ and
        U. Heiter$^{4}$ \\
        $^1$Univ. Bordeaux, LAB, UMR 5804, F-33270, Floirac, France. \\
        $^2$CNRS, LAB, UMR 5804, F-33270, Floirac, France. \\
        $^3$Institute of Astronomy, University of Cambridge, Madingley Road, Cambridge CB3 0HA, U.K. \\
        $^4$Department of Physics and Astronomy,  Uppsala University, Box 516, 75120 Uppsala, Sweden. \\
        }

\pubyear{2014}
\volume{10}
\pagerange{\pageref{firstpage}--\pageref{lastpage}}

\date{Received --- ; accepted ---}

\maketitle
\label{firstpage}

\begin{abstract}
iSpec is an integrated spectroscopic software framework suitable for the creation of spectral libraries such as the Benchmark Stars library and the determination of atmospheric parameters (i.e. effective temperature, surface gravity, metallicity) and individual chemical abundances.

The framework can be used in automatic massive analysis, but it also includes a user-friendly visual interface that can easily interoperate with other astronomical applications such as TOPCAT, VOSpec and splat which provide access to the Virtual Observatory.

\end{abstract}


\section{Introduction}\label{s:intro}

On-going high resolution spectroscopic surveys such as the Gaia ESO survey \citep[GES,][]{2012Msngr.147...25G} and the future HERMES/GALAH \citep{2010gama.conf..319F} provide an enormous quantity of high quality spectra. This represents a unique opportunity to unravel the history of our Galaxy by studying the chemical signatures of large samples of stars. 

This huge quantity of spectroscopic data challenges us to develop automatic processes to perform the required analysis. Numerous automatic methods have been developed over the past years \citep[e.g][to name a few]{1996A&AS..118..595V, 1998A&A...338..151K, 2006MNRAS.370..141R, 2009A&A...501.1269K, 2010A&A...517A..57J, 2013ApJ...766...78M, 2013A&A...558A..38M} to treat the spectra and derive atmospheric parameters from large datasets, where each of them uses different reference models and minimization strategies. A consequence of this diversity is the inhomogeneity of the published measurements. This can be seen when we compare atmospheric parameters from different sources in the bibliographical compilation PASTEL \citep{2010A&A...515A.111S}.

To improve the understanding of the systematics between the various characterization methods, we have assembled a library of the 34 Gaia FGK Benchmark Stars, covering different regions of the HR diagram and spanning a wide range in metallicity (see \citealt{IWSSL2013Proceedings} in this proceedings). Their temperature and surface gravity were determined from fundamental relations and [Fe/H] from a dedicated spectroscopic analysis \citep{2013arXiv1309.1099J}.

In Sect.~\ref{s:framework}, we describe the iSpec software framework\footnote{We plan to freely distribute iSpec under the Open Source terms of the GNU Affero General Public License.} that was used to prepare the spectra of the library (Blanco-Cuaresma et al, submitted). In Sect.~\ref{s:chemical_tagging}, we show an application of iSpec to chemical tagging by using its capabilities for the determination of atmospheric parameters and abundances (Blanco-Cuaresma et al, in preparation).




\section{The spectroscopic software framework}\label{s:framework}

The main functionalities for spectra treatment and library creation that are integrated into iSpec are the following:

\begin{itemize}
    \item \textbf{Cosmic rays removal:} Spectra may contain residuals from cosmics rays which are automatically detected by using a median filter to smooth out single flux deviations.
    \item \textbf{Continuum normalization:} The continuum points of a spectrum are found by applying a median and maximum filter with different window sizes. Afterwards, a polynomial or group of splines can be use to model the continuum and finally normalize the spectrum by dividing all the fluxes by the model.
    \item \textbf{Resolution degradation:} The spectral resolution can be degraded by convolving the fluxes with a Gaussian of a given FWHM (km/s).
    \item \textbf{Radial velocity:} By cross-correlating \citep{2007AJ....134.1843A} the spectrum with a mask or template (e.g. observed or synthetic), the radial velocity of the star can be derived and corrected.
    \item \textbf{Telluric lines identification:} Telluric lines from Earth's atmosphere contaminate ground-observed spectra and it can affect the parameter determination. Their position in the spectra can be determined by cross-correlating with a telluric mask build from a synthetic spectrum (thanks to TAPAS \citealt{2013arXiv1311.4169B}).
    \item \textbf{Resampling:} Spectra can be re-sampled by using linear (2 points) or Bessel (4 points) interpolation \citep{1998A&A...338..151K}.
\end{itemize}

Additionally, iSpec is capable of determining atmospheric parameters (i.e effective temperature, surface gravity, metallicity, micro/macroturbulence, rotation) and individual chemical abundances. Two different techniques can be applied:

\begin{itemize}
    \item \textbf{Synthetic spectra fitting technique:} iSpec compares an observed spectrum with synthetic ones generated on the fly, in a similar way as Spectroscopy Made Easy (SME) tool does \citep{1996A&AS..118..595V}. A least-squares algorithm (Levenberg-Marquardt implementation) minimizes the difference between the synthetic and the observed spectra in order to converge towards the closest synthetic spectrum. In each iteration, the algorithm will vary each one of the free parameters at a time and prognosticate in what direction it should move to find the solution. Specific regions of the spectrum can be selected to minimize the computation time and focus onto the most relevant parts to better identify stars (i.e. wings of H-$\alpha$/MgI triplet, Fe I/II lines).
    \item \textbf{Equivalent widths method:} Gaussian models are fitted for a given list of Fe I/II lines, from their integrated area we can derive their respective equivalent width and thus their abundances. The algorithm for determining the atmospheric parameters is a least-squares technique (as explained above) based on the assumption of excitation and ionization equilibrium, similar to GALA \citep{2013ApJ...766...78M} and FAMA \citep{2013A&A...558A..38M}.
\end{itemize}

``SPECTRUM: a Stellar Spectral Synthesis Program'' \citep{1994AJ....107..742G} is used to generate synthetic spectra and derive abundances from equivalent widths. iSpec offers an wide range of options such as different atomic line lists, model atmospheres (i.e. ATLAS \citealt{2005MSAIS...8...14K} or MARCS \citealt{2008A&A...486..951G}) and solar abundances.

It is worth noting that the framework can be used in automatic massive analysis through python scripts, but it also includes a user-friendly visual interface that can easily interoperate with other astronomical applications such as TOPCAT\footnote{\url{http://www.star.bris.ac.uk/\string~mbt/topcat/}}, VOSpec\footnote{\url{http://www.sciops.esa.int/index.php?project=ESAVO\&page=vospec}} and splat\footnote{\url{http://star-www.dur.ac.uk/\string~pdraper/splat/splat.html}} facilitating a indirect way to access the Virtual Observatory\footnote{\url{http://www.ivoa.net/}}

\begin{figure}
    \begin{centering}
        \includegraphics[scale=0.50,trim = 1mm 1mm 1mm 1mm, clip]{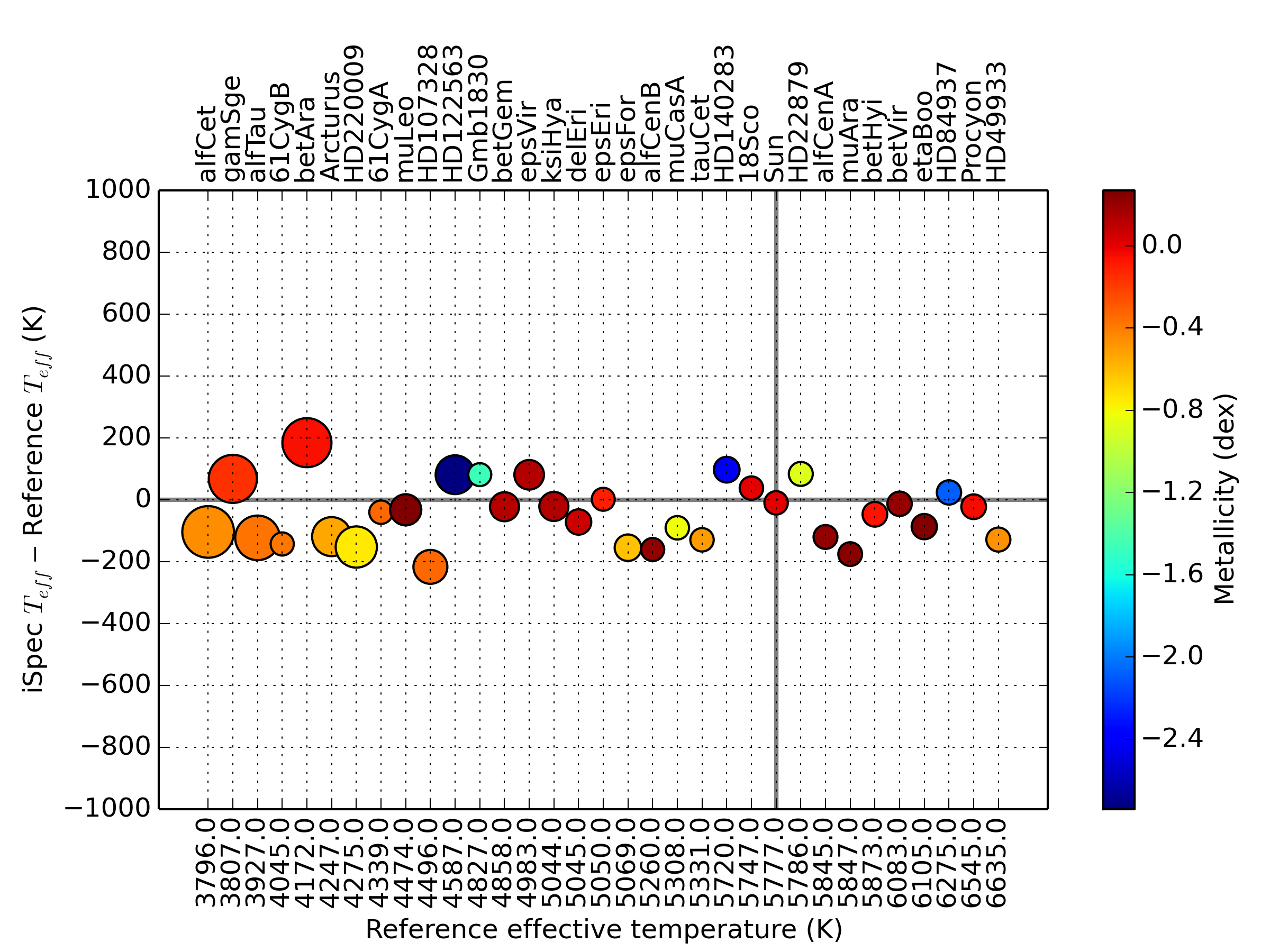}
        \par
    \end{centering}
    \caption{Differences in effective temperature between the reference (the Gaia FGK Benchmark Stars) and the derived value by iSpec (synthetic spectral fitting method). Stars are sorted by temperature, the color represents the metallicity and the size is linked to the surface gravity.}
    \label{fig:bs_teff}
\end{figure}

\begin{figure}
    \begin{centering}
        \includegraphics[scale=0.50,trim = 1mm 1mm 1mm 1mm, clip]{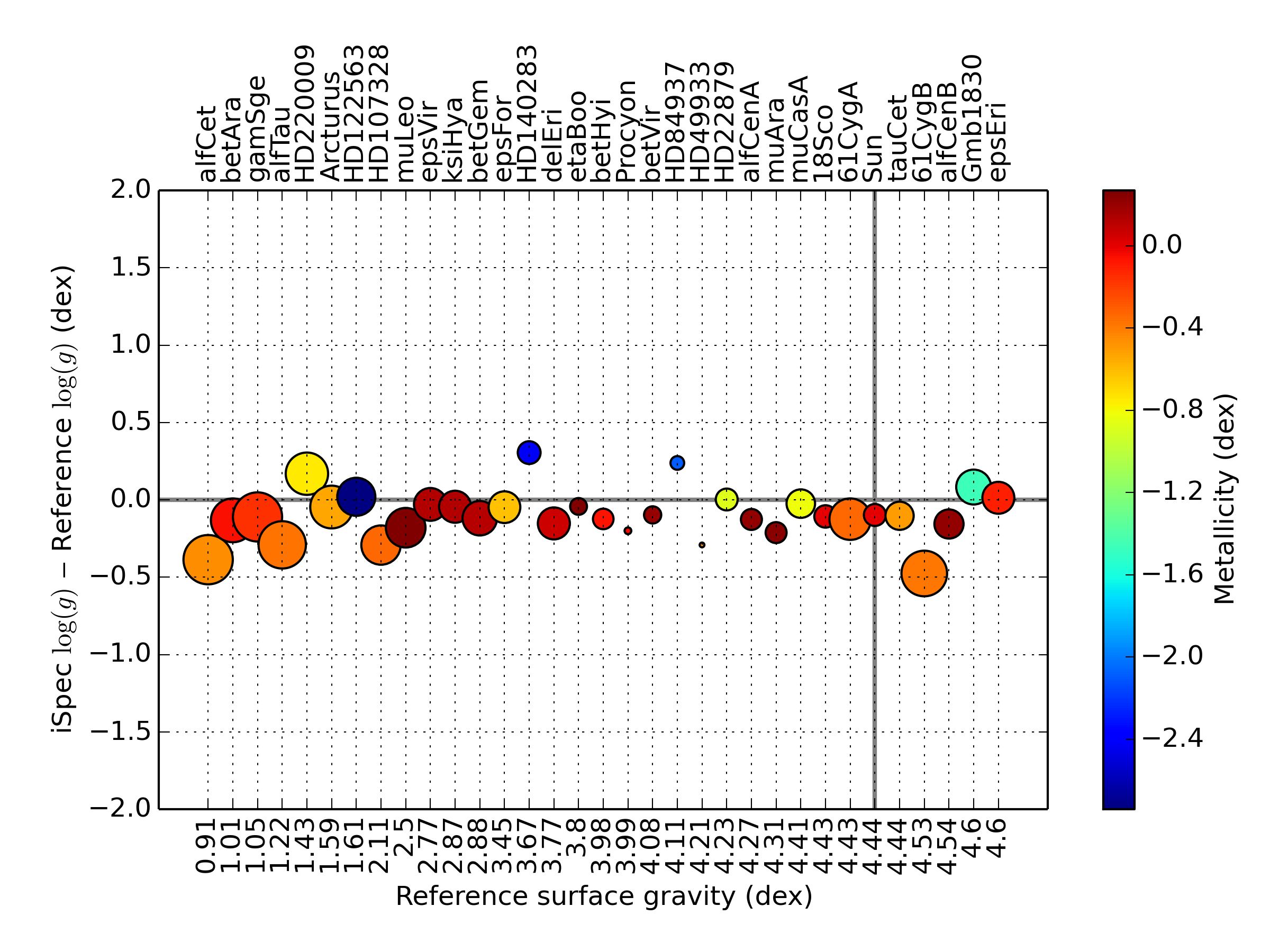}
        \par
    \end{centering}
    \caption{Differences in surface gravity between the reference (the Gaia FGK Benchmark Stars) and the derived value by iSpec (synthetic spectral fitting method). Stars are sorted by surface gravity, the color represents the metallicity and the size is linked to the temperature.}
    \label{fig:bs_logg}
\end{figure}

\begin{figure}
    \begin{centering}
        \includegraphics[scale=0.50,trim = 1mm 1mm 1mm 1mm, clip]{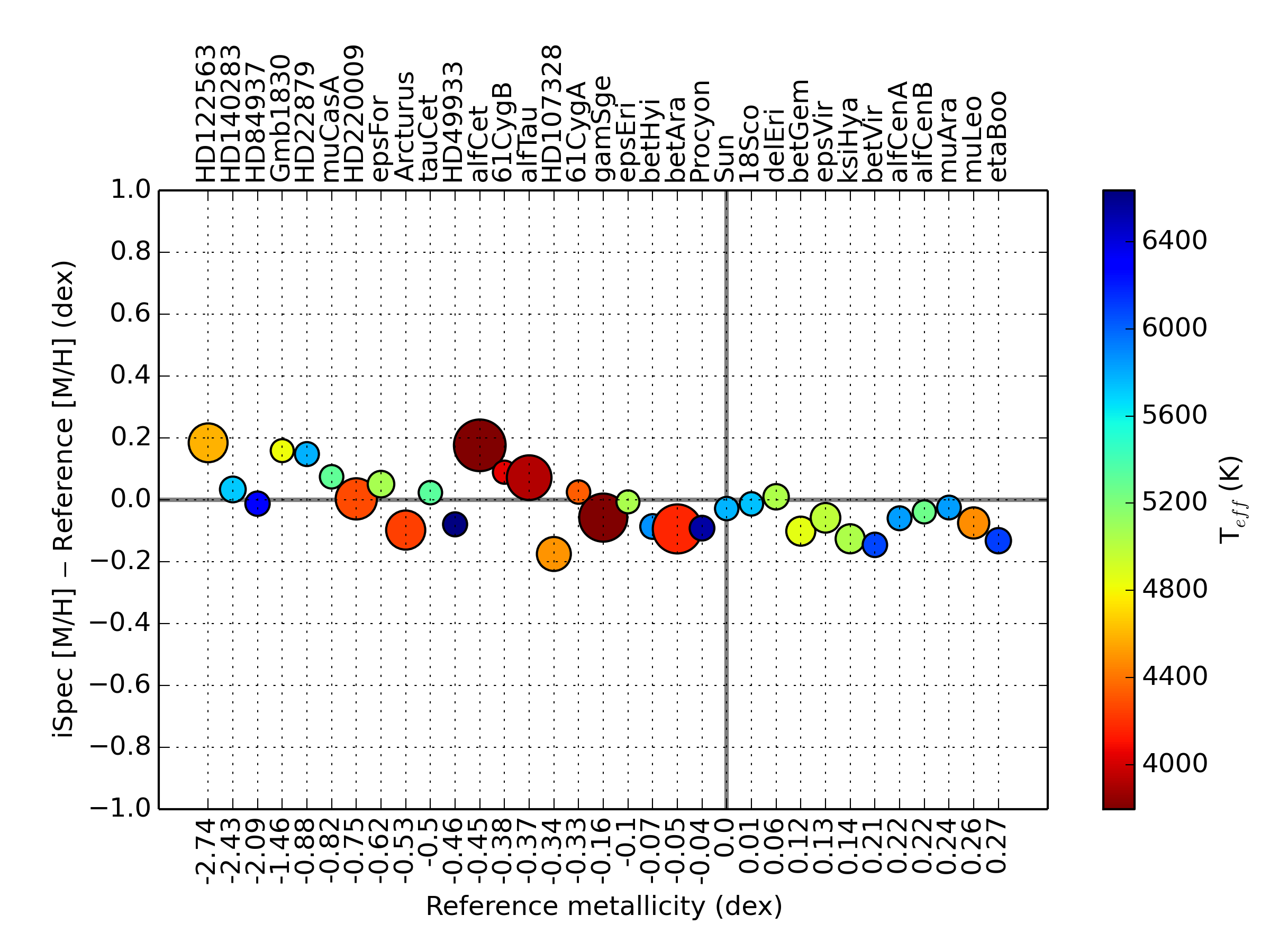}
        \par
    \end{centering}
    \caption{Differences in metallicity between the reference (the Gaia FGK Benchmark Stars) and the derived value by iSpec (synthetic spectral fitting method). Stars are sorted by metallicity, the color represents the temperature and the size is linked to the surface gravity.}
    \label{fig:bs_MH}
\end{figure}

The derivation of atmospheric parameters has been tested using the Gaia FGK Benchmark Stars library and associated parameters (see Sect.~\ref{s:intro}). The results are very consistent as shown in Fig.~\ref{fig:bs_teff}, \ref{fig:bs_logg} and \ref{fig:bs_MH}.

\section{Chemical tagging}\label{s:chemical_tagging}

Stars born from the same giant molecular cloud are expected to share the same homogeneous chemical patterns as shown by \cite{2007AJ....133.1161D}. These findings open the possibility of using the technique of chemical tagging to identify common formation sites in the disk as proposed by \cite{2002ARA&A..40..487F}. This requires to apply automatic and massive spectral analysis to handle huge quantities of spectra.

\begin{figure}[H]
    \begin{centering}
        \includegraphics[scale=0.50, trim = 1mm 1mm 1mm 1mm, clip]{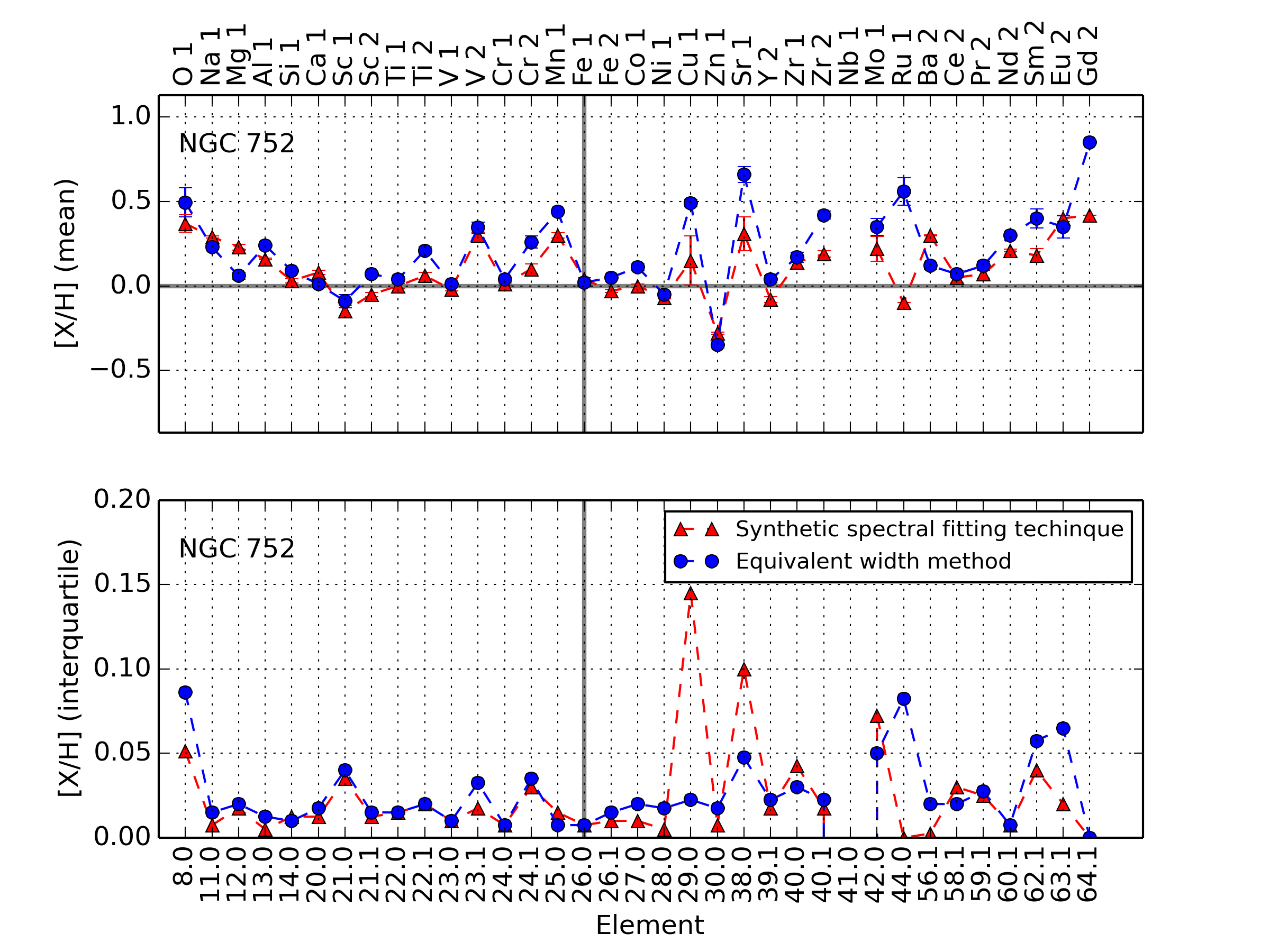}
        \par
    \end{centering}
    \caption{Chemical abundances (upper) and dispersion (lower) of the open cluster NGC 752.}
    \label{fig:chemical_tagging_1}
\end{figure}

\begin{figure}[H]
    \begin{centering}
        \includegraphics[scale=0.50, trim = 1mm 1mm 1mm 1mm, clip]{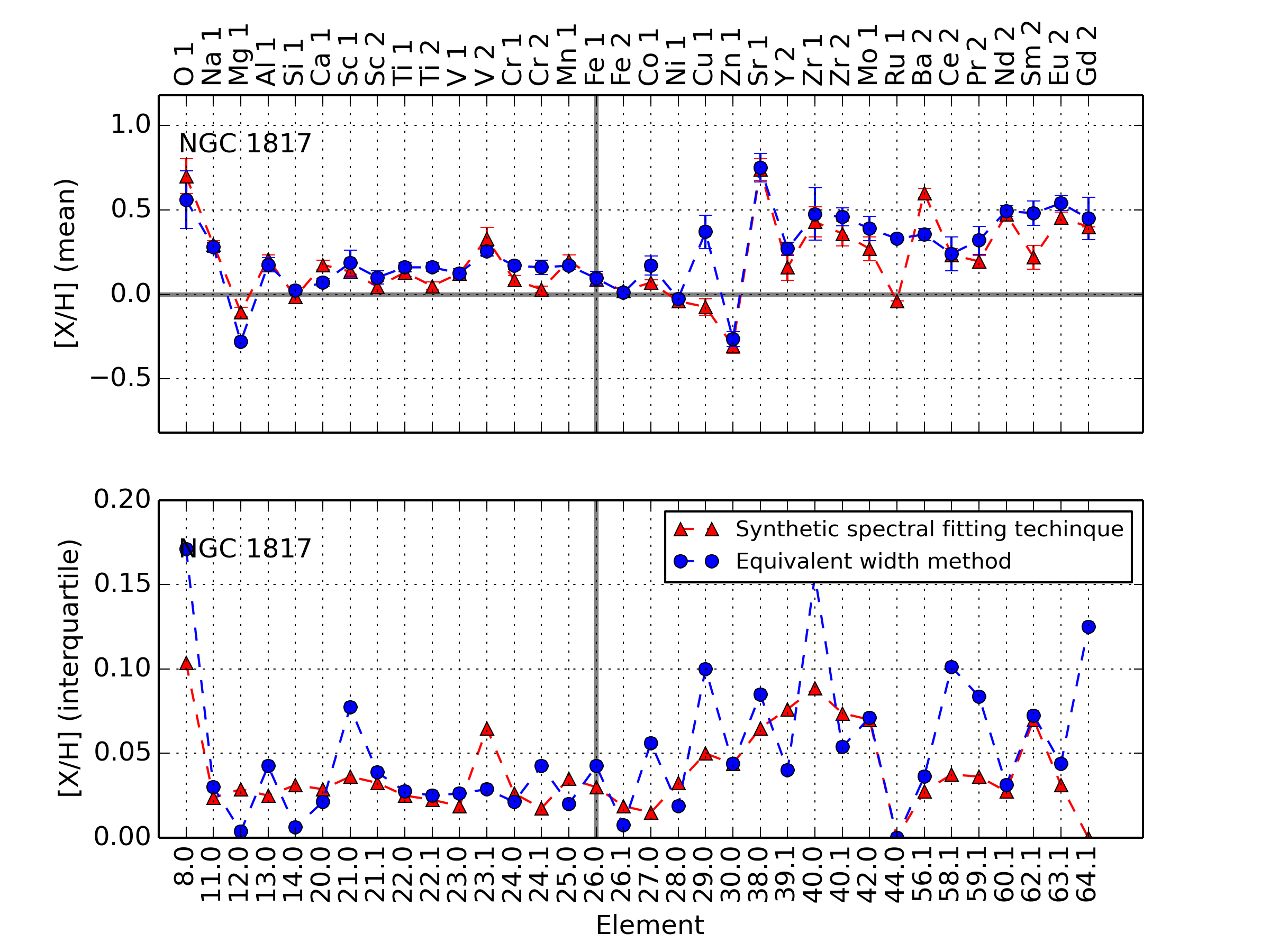}
        \par
    \end{centering}
    \caption{Chemical abundances (upper) and dispersion (lower) of the open cluster NGC 1817.}
    \label{fig:chemical_tagging_2}
\end{figure}

\begin{figure}[H]
    \begin{centering}
        \includegraphics[scale=0.50, trim = 1mm 1mm 1mm 1mm, clip]{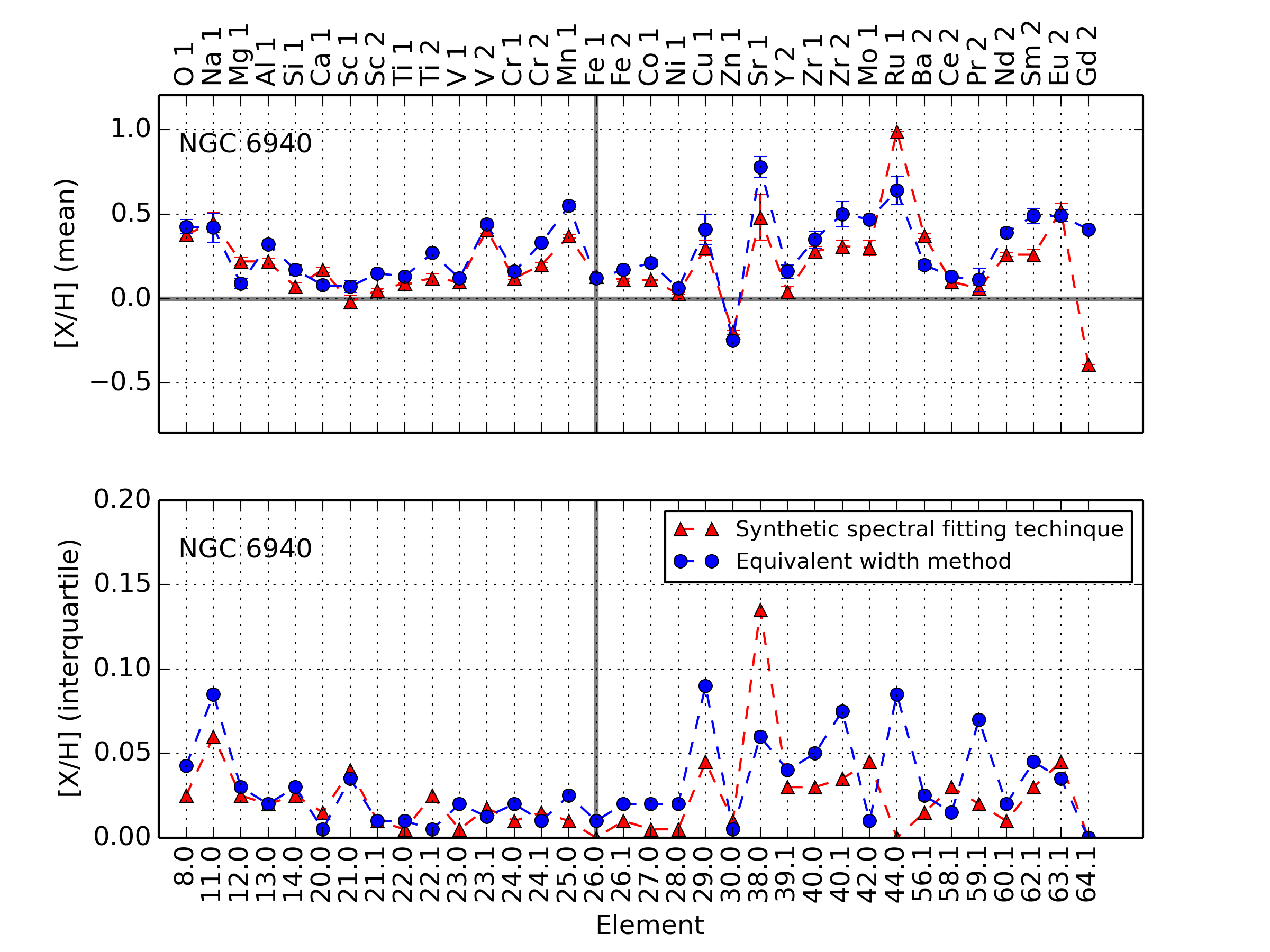}
        \par
    \end{centering}
    \caption{Chemical abundances (upper) and dispersion (lower) of the open cluster NGC 6940.}
    \label{fig:chemical_tagging_3}
\end{figure}

We have tested and validated the performance of iSpec for this purpose by automatically analy\-sing three different open clusters observed by the NARVAL spectropolarimeter mounted on the 2m Telescope Bernard Lyot \citep{2003EAS.....9..105A} (located at Pic du Midi, France). After deriving their atmospheric parameters and the individual abundances for a list of given elements, we have shown that stars belonging to the cluster share a common chemical pattern, with a very low dispersion for certain elements (below 0.05~dex). Additionally, the synthetic spectra fitting technique and equivalent width method are able to recover similar chemical patterns as shown in Fig.~\ref{fig:chemical_tagging_1}, \ref{fig:chemical_tagging_2} and \ref{fig:chemical_tagging_3}. A detailed discussion of the abundances of individual elements will be presented in a forthcoming paper.

\section{Conclusions}\label{s:conclusions}

iSpec has been proven to be a suitable framework for the creation of spectral libraries such as the Gaia FGK Benchmark Stars (Blanco-Cuaresma et al., submitted), together with the derivation of atmospheric parameters and individual chemical abundances (Blanco-Cuaresma et al., in preparation). The code will be released on \url{http://www.blancocuaresma.com}



\clearpage

\section*{Acknowledgements}

This work was partially supported by the Gaia Research for European Astronomy Training (GREAT-ITN) Marie Curie network, funded through the European Union 7th Framework Programme [FP7/2007-2013] under grant agreement n. 264895.
All the software used in the data analysis has been provided by the Open Source community.


\label{lastpage}
\end{document}